\documentclass[12pt,preprint]{aastex}

\shorttitle{McNeil's Nebula in Orion: The Outburst History}
\shortauthors{Brice\~no et al.}

\begin{document}

\title{Mc Neil's Nebula in Orion: The Outburst History\altaffilmark{1}}


\author{C. Brice\~no\altaffilmark{2}, A.K. Vivas, J. Hern\'andez\altaffilmark{3}}
\affil{Centro de Investigaciones de Astronom{\'\i}a (CIDA),
    Apartado Postal 264, M\'erida 5101-A, Venezuela}
\email{briceno@cida.ve}

\author{N. Calvet, L. Hartmann, T. Megeath, P. Berlind and M. Calkins}
\affil{Smithsonian Astrophysical Observatory, 60 Garden St, Cambridge, MA 02138, USA}

\and

\author{S. Hoyer\altaffilmark{4}}
\affil{Dept. Astronom{\'\i}a y Astrof{\'\i}sica,
	Pontificia Universidad Cat\'olica de Chile,
 	Campus San Joaquin, Vicu\~na Mackenna 4860 Casilla 306
	Santiago 22, Chile}

\altaffiltext{1}{Based on observations obtained at the
Llano del Hato National Astronomical Observatory of Venezuela,
operated by CIDA for the Ministerio de Ciencia y Tecnolog{\'\i}a,
and at the Fred Lawrence Whipple Observatory of the Smithsonian
Institution, USA}
\altaffiltext{2}{Research Associate at SAO, 60 Garden St., Cambridge, MA 02138, USA}
\altaffiltext{3}{Postgrado en F{\'\i}sica Fundamental, Universidad de los Andes,
M\'erida, Venezuela}
\altaffiltext{4}{Summer Student at CIDA, M\'erida, Venezuela}

\begin{abstract}
We present a sequence of I-band images obtained at the Venezuela 
1m Schmidt telescope during the outburst of the nebula recently 
discovered by J.W. McNeil in the Orion L1630 molecular cloud. 
We derive photometry spanning the pre-outburst state and the 
brightening itself, a unique record including 14 epochs and spanning 
a time scale of $\sim 5$ years. We constrain the beginning of the 
outburst at some time between Oct. 28 and Nov. 15, 2003.
The light curve of the object at the vertex of the nebula, the likely 
exciting source of the outburst, reveals that it has brightened $\sim 5$ 
magnitudes in about 4 months. The time scale for the nebula to develop 
is consistent with the light travel time, indicating that we are observing 
light from the central source scattered by the ambient cloud into the line 
of sight.
We also show recent FLWO optical spectroscopy of the exciting source 
and of the nearby HH 22. The spectrum of the source is highly reddened; 
in contrast, the spectrum of HH 22 shows a shock spectrum superimposed 
on a continuum, most likely due to reflected light from the exciting 
source reaching the HH object through a much less reddened path.
The blue portion of this spectrum is consistent with an early B spectral 
type, similar to the early outburst spectrum of the FU Ori variable 
V1057 Cyg; we estimate a luminosity of ${\rm L \sim 219 L_\sun}$.
The eruptive behavior of the McNeil nebula source, its spectroscopic
characteristics and luminosity, suggest we may be witnessing an FU Ori 
event on its way to maximum. Further monitoring of this object will 
decide whether it qualifies as a member of this rare class of objects.

\end{abstract}

\keywords{stars: formation ---
stars: pre-main sequence --- stars: variables: other
---  ISM: Herbig-Haro objects}

\section{Introduction}

Last February 9, 2004, the discovery of a
new nebula, roughly 12' south of the reflection nebula
NGC 2068 (M78) in the L1630 molecular cloud of the Orion star-forming complex 
was announced by \cite{mcn04}. 
The region surrounding the newly revealed object contains a number
of pre-main sequence stars, as well as Herbig-Haro objects HH 22 and HH 23
\citep{eim97}.
The IRAS source 05436-0007 \citep{cla91} is located at
the vertex of the new nebula, but was not identified by him 
as a young object. The 2MASS images show an extremely
red source at this position (J05461313-0006048),
but no object is visible at optical
wavelengths in Palomar Observatory Sky Survey plates.
More recently, dust continuum imaging of this region by \citet{lmz99}
revealed several sources at 350 and $1300\mu$m. They suggest that their
source LMZ 12, spatially coincident with IRAS 05436-0007, is a Class 0 source 
with $L_{bol}\simeq 2.7 L_\sun$,
and the probable
exciting source of HH 23, located $\sim 2\farcm 5$ north of the new nebula.

This new nebula was made widely
known by B. Reipurth's announcement in the Star Formation
Newsletter No. 136\footnote{http://www.ifa.hawaii.edu/$\sim$reipurth/newsletter.htm}. 
In a follow up study \citet{rea04} indicate that the object
brightened by $\sim 3$ magnitudes in the near infrared, and that
its optical and K-band spectra show a certain resemblance with EXors.
On the other hand, the object may be 
undergoing an FU Ori type eruption, which would imply
a larger variation in brightness
and a longer period at maximum light (Herbig 1977). To discriminate
between these possibilities both pre-outburst and post-outburst 
observations are needed.

We report here optical images, photometry
and spectroscopy of the McNeil Nebula,
previous to its brightening, and during the outburst itself.
These observations span $\sim$ 5 years, 
with a particularly good coverage
of the time when the outburst started.
The observations were obtained during
a sensitive optical variability study 
spanning most of the Orion OB1 Association (Brice\~no et al. 2001).  
  
\section{Observations and Data Analysis}

Our large scale photometric variability study of the Orion
OB1 association (see Brice\~no et al. [2001] for details) 
has been carried out with the QuEST camera
on the 1m Schmidt telescope at the Llano del Hato National Astronomical
Observatory in Venezuela
(see Baltay et al. [2002] for details of the instrument).
The camera consists of 16 (in a $4 \times 4$ array) front illuminated 
$2048 \times 2048$ CCDs with $15\mu$ pixels,
although only three rows of 4 CCDs are presently operational; 
the plate scale is $1.03 \arcsec$/pix. 
The instrument is optimized to work in drift-scan mode along
strips of the sky, $2\fdg 3$ wide, at constant declination.
Each row of 4 CCDs is fitted with a different filter. 
The equivalent exposure time (the time an
object takes to cross a chip at $\delta = 0\arcdeg$) is 140s.

The region we have been monitoring since January 1999 is centered at 
$\delta=+1\fdg 0$, and spans the range of $\alpha=5^h 10^m$ to
$5^h 48^m$. We used one V and two I filters to cover the three good rows
of CCDs. Given the location of the source of McNeil's 
nebula ($\alpha_{J2000}=5^h 46^m 13\fs 1; 
\delta_{J2000}=-0\arcdeg 6\arcmin 5\arcsec$), 
and because the V filter is located 
in the westernmost part of the array, we detected the nebula
only in the I band at the very end of most scans. 

We were able to recover the nebula in 19 observations on 14 different nights,
from January 9, 1999 to January 28, 2004 (Table 1). We obtained
an additional I-band observation on February 26, 2004 with the 4Shooter
CCD Mosaic camera installed on the SAO 1.2m telescope on Mt. Hopkins, Arizona.
Thus, our dataset spans 1874 days
and includes the period when the nebula began to brighten up.
All observations were obtained at an airmass of $\sim 1$, and
the seeing ranged from $\sim 2 - 3.5$ arcsec. 
The reduction and processing of the Schmidt data was done with
custom made automated software (see Vivas et al. [2004] for details)
that produces magnitudes and accurate positions for all point sources 
in the region, including a variability flag. 
The 4Shooter images were processed using custom IDL routines.

In Fig. 1 we show a sequence of I-band images obtained between Oct.25, 2003
and Jan.26, 2004 at Llano del Hato.
A faint source is barely visible in Oct. 2003,
and it begins to brighten up significantly between Nov 15 and 23, 2003,
when the nebula begins to be barely detectable to the NW.
The nebula exhibits
significant brightening and evolution from this date on,
while the source at its vertex continues to get brighter.
By Jan.20, 2004
the object had mostly reached the state it revealed in the discovery
and subsequent images.

In order to measure the brightness of the
new nebula and its apparent source, we selected 14 reference stars
calibrated as secondary standards with \citet{lan92} fields,
with magnitudes ${\rm I_C}\sim$=12 - 16 and colors V-${\rm I_C}\sim$ 1.4 - 3.9;
these stars are located near the nebula ($\la 10\arcmin$) and have on average 25
measurements spanning 5 years, showing no significant 
variability over that time frame, thus
making them robust comparison stars. 
We could not use our automated software for the new observations of the
nebula because it does not handle photometry of extended objects. 
Thus, we used standard IRAF\footnote{IRAF is distributed by the National Optical Astronomy Observatories,
    which are operated by the Association of Universities for Research
    in Astronomy, Inc., under cooperative agreement with the National
    Science Foundation.}
routines to perform differential photometry
in several parts of the nebula.
We measured three different positions within the
nebula: A, B and C, shown in Fig. 2; these
were defined as offsets from the star labeled $V$ in Fig. 2 
($\alpha_{J2000}=5^h 46^m 18\fs 9;
\delta_{J2000}=-0\arcdeg 5\arcmin 38\arcsec$).
In this way, we were able to
measure early images when the nebula was barely detected or not seen at all.
Because our images had FWHM $\sim 3\arcsec$, we used an aperture radius=$4\farcs 1$
(for comparison stars and locations on the nebula).
The region we call A is centered on the nebula vertex,
located $1\arcmin 26\farcs 1$ W, and $26\farcs 2$ S 
from star $V$.
In all images this object looks like a point source.\footnote{
Because of the large aperture our measurements include 
nebulosity inmediately surrounding this object. Aspin (2004),
private communication,
measured the source at the vertex through a much smaller aperture ($0.9 \arcsec$)
on Gemini North GMOS images taken on UT Feb. 03, 2004 through a SDSS i' filter,
and found a 1.2 magnitude difference with respect to our Feb. 26, 2004 value.
}
Our second aperture (B) was centered on the nebulosity just north of the star.
Its location is $1\arcmin 28\farcs 6$ W, $18\farcs 1$ S from $V$.
Finally, the third 
aperture (C) was placed on top of the object HH 22, which also brightened
during the event (see Fig. 1), centered at $1\arcmin 12\farcs 1$ W, 
$7\farcs 7$ N of $V$. 
Since the
region immediately surrounding the apertures is filled with nebulosity we could not 
use an annulus to measure the sky brightness. Instead, the sky was measured
by taking the average value of small boxes (3x3 pixels) in 10-12
different locations outside and near the nebula. 

In Fig. 3 we show the resulting light curve of A, B, and C.
Our data suggest that even in its "quiescent" phase, with a 
mean magnitude $\rm{I_C}= 19.1$, source A is variable on time 
scales of a few years at the 0.7 mag level, with an extreme range 
of $\la 2$ mags (the nebular structure is not visible in 
our pre-outburst images);
this is consistent with reports of this object appearing
in an image from 1966, 
although rather fainter than at present \citep{skt04}.
By the time of our first 2003 observations on Oct. 25 and 28,
source A had brightened by $\sim 0.8$ mags compared to Jan. 2003.
But the more dramatic rise in brightness
occurred sometime after Oct. 28 and before Nov. 15, when it was 
1.8 mag brighter than its mean level.
We measured a total increase of $\sim 5$ magnitudes in about 4 months.
Source B closely traces the behavior of A after Oct. 2003,
but is on average 0.6 mags 
fainter and its brightness
does not increase significantly until Nov. 27, a delay of
roughly 18 days respect to A.
Source C, spatially coincident with HH 22, is close to our detection
threshold of $\rm{I_C}\sim 19$ until Dec.15, when it is clearly
visible at $\rm{I_C}= 18.2$. This represents a delay of $\ga 50$ 
days with respect to source A.
Finally, our most recent data from Feb.26, 2004, strongly
suggests that the rate of brightening has diminished.
If we assume a distance
of 400 pc as for HH 22 and 23 \citep{ant82}, and consider the separations
A-B= $8\farcs 5$ (0.016 pc) and A-C= $36\farcs 7$ (0.071 pc), we derive
a light travel time of 19 days between A and B, and 85 days between A and C.
This is consistent with the delays inferred from the
light curve in Fig.2, and indicates that we are observing light
scattered into the ambient cloud, possibly coming out through a cavity
opened in the embedded source's envelope.

Optical spectra were obtained the night of Feb.18. 2004
at positions A and C of the new nebula,
using the 1.5 meter telescope of the
Whipple Observatory with the FAST Spectrograph \citep{fab98}, equipped with the Loral
$512 \times 2688$ CCD. 
The spectrograph was set up with a
300 groove $mm^{-1}$ grating and $5\arcsec$
wide slit (oriented at PA = 90$^\circ$),
yielding 3400 {\AA } of spectral coverage centered
at 5500 {\AA} and a resolution of $\sim 10${\AA }.
The spectra were reduced and wavelength calibrated 
using the standard IRAF routines.
The effective exposure time at each slit position was 1200s.
The spectra were corrected for the relative system response 
using the IRAF {\it sensfunc} task and
observations of spectrophotometric standard stars.

Spectra of A and C are shown in the
upper panel of Fig. 4.
The spectrum of A corresponds to a heavily
reddened source, in agreement with the $A_V \ge 12$
derived from 2MASS JHK$_s$ colors;
depending upon the spectral type (see below)
and reddening law adopted, the continuum in the visual
range suggests $A_V \sim 8-10$, which may be underestimated
since there may be a significant scattered light component.
The most conspicuous feature is strong emission
at H$\alpha$. In contrast, the spectrum of C
is much flatter, indicating much less
reddening; it also exhibits a number of forbidden lines,
in particular [OI] 6300, 6363{\AA};
[SII] 6716, 6730 {\AA}; 
[NII] 6548, 6583 {\AA};
and emission in H$\alpha$ and H$\beta$. 
The blue portion of the spectrum of C shows
the higher Balmer lines in absorption; in the lower
panel of Fig. 4, we compare 
with spectroscopic standards. 
The strength of the Balmer lines is consistent with
either an early B or an early F type, as shown
in the figure. However, the absence of the G band
and the weakness of Ca II K favor an early B spectrum. 

We suggest that we have here a situation
similar to the case of the embedded FU Ori object L1551,
where the optical spectrum of the object
was obtained by observing its light reflected by the
nearby HH 102 (Mundt et al. 1985).
If the line of sight
from A to C has been cleared out by the outflow
which created HH 22 at the wind-cloud interaction region,
then the spectrum of A that reaches C is much less reddened
than that in Earth's direction.
The spectrum of C consists of
an emission line spectrum, arising mostly in situ
from the shock front(s), plus the reflected, lightly reddened spectrum
of A. Thus, the spectrum of C gives us direct information of the
spectrum of the driving source, which in the blue appears
to be close to that of an early B star.

\section{Discussion}
 
The behavior of the McNeil nebula is reminiscent of the early evolution
of the outbursts of FU Ori and V1057 Cyg (Herbig 1977).
The optical (red) rise of McNeil, $\sim 5$ mag in about a third
of a year, is not much different than the early rise of these two objects,
prototypical of the FUor class;
moreover, the rise time of another FUor, V1515 Cyg, was much longer,
indicating that the rate of brightening is not a universal quality
among these objects. Like the McNeil object, V1057 Cyg initially
showed a P Cygni profile on strong H$\alpha$ emission
(see inset in Figure 4), evolving into stronger (blue-shifted)
absorption later on. Strong absorption in the upper Balmer lines was
also present in spectra of V1057 Cyg, classified as
B3 by \citep{wel71}. However, the lack of He I lines lead Herbig to assign
an approximate type of early A (Herbig 1977).  Herbig also noted the
presence of wide emission at the Ca II resonance lines, which lead to
a very weak Ca II 3933 {\AA} absorption; this is also consistent with
our optical spectra.

The eruptive behavior suggests that this may be an FU Ori object
on its way to maximum, or possibly a member of the so-called class of
EXors (Herbig 1977), which exhibit similar or smaller increases in optical
brightness but last shorter periods of time.
Continued monitoring should better help distinguish between these possibilities.
Given the small number of objects in the FUor and EXor classes, it may
be that there exists a continuum of outburst behavior which will
be filled in by the discovery of additional objects.
We note that at its recent brightness of ${\rm I_c} = 14.4$, assuming 
${\rm A_I} = 7.2$,
a distance of 400 pc, and adopting an A0 spectral type 
as a compromise between the range of spectral types we inferred before,
the system luminosity
is $L \sim 219 {\rm L_\sun}$ (and would be higher if the intrinsic spectrum is
earlier); this large luminosity is more similar to FUors than EXors.
 
The eruptive mechanism of FUors, and likely also that of EXors, is
thought to be rapid outbursts in disk accretion (Hartmann \& Kenyon 1996),
possibly driven by the pile up of material in a disk due to rapid
protostellar envelope infall (Kenyon \& Hartmann 1991).
The estimated bolometric luminosity  $L_{bol}\simeq 2.7 L_\sun$
and a SED like that of a Class 0 source \citep{lmz99}
is consistent with the pre-outburst object being a low-mass protostar.
This object looks more similar to FU Oris, which are often highly
embedded (Hartmann \& Kenyon 1996) than EXors, which
seem to be much less extincted.  This very interesting source warrants
further monitoring to help determine its true nature.
 
\acknowledgments

We are grateful to C. Castillo, J.J. Downes, M. Mart{\'\i}nez, 
F. Molina, L. Torres, 
O. Contreras, F. Moreno, G. Rojas, and U. S\'anchez for 
obtaining the observations for this article. We also acknowledge
the prompt review and helpful comments from the referee Dr. Colin Aspin.
C. Brice\~no acknowledges support from grant S1-2001001144 of
FONACIT, Venezuela.
Part of this work was supported by 
NSF grant AST-9987367 and NASA grant NAG5-10545.

Facilities: \facility{NOV(Schmidt+QUEST)}, \facility{FLWO(1.2m+4shooter)}, \facility{FLWO(1.5m+FAST)}.

\clearpage

\begin{figure}
\epsscale{1.1}
\plotone{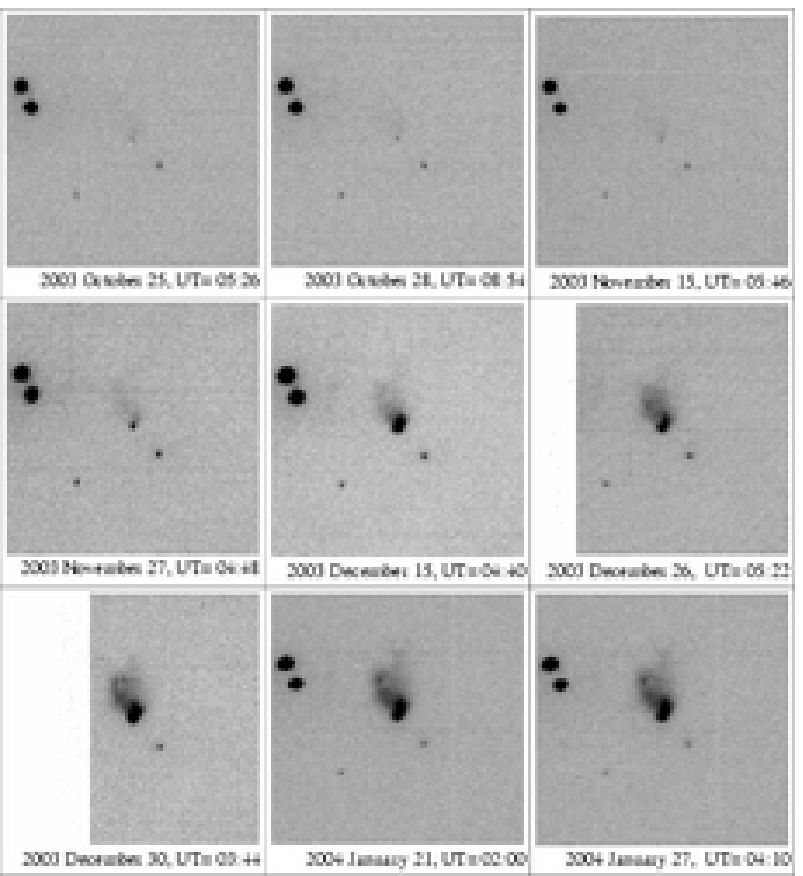}
\caption{Sequence of the outburst in the I-band. Each frame is 
$3.5\arcmin \times 3.5\arcmin$. North is up, East is left. }
\end{figure}

\begin{figure}
\rotate
\epsscale{1.00}
\plotone{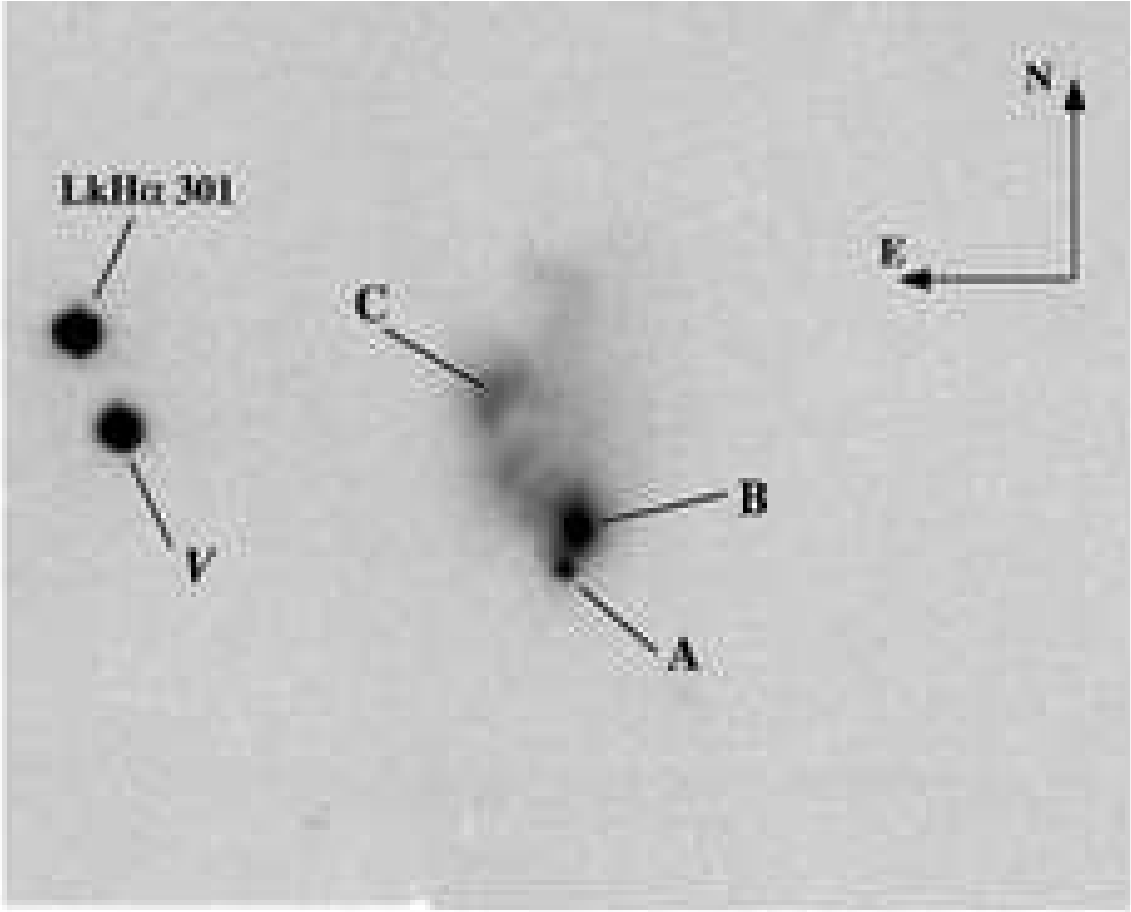}
\caption{R-band exposure (5min) obtained in point-and-stare
mode with the Venezuela Schmidt telescope on Feb.16, 2004
(UT=02:48:27).
The field is 3.5' $\times$ 3.0'. We show the three features
in the nebula which were measured in our I-band imaging data.
We indicate the known T Tauri star LkH$\alpha$ 301. Star $V$ was
found to be variable in our data, and our recent FAST spectra
of this object confirm it is a T Tauri star.}
\end{figure}

\begin{figure}
\epsscale{0.9}
\plotone{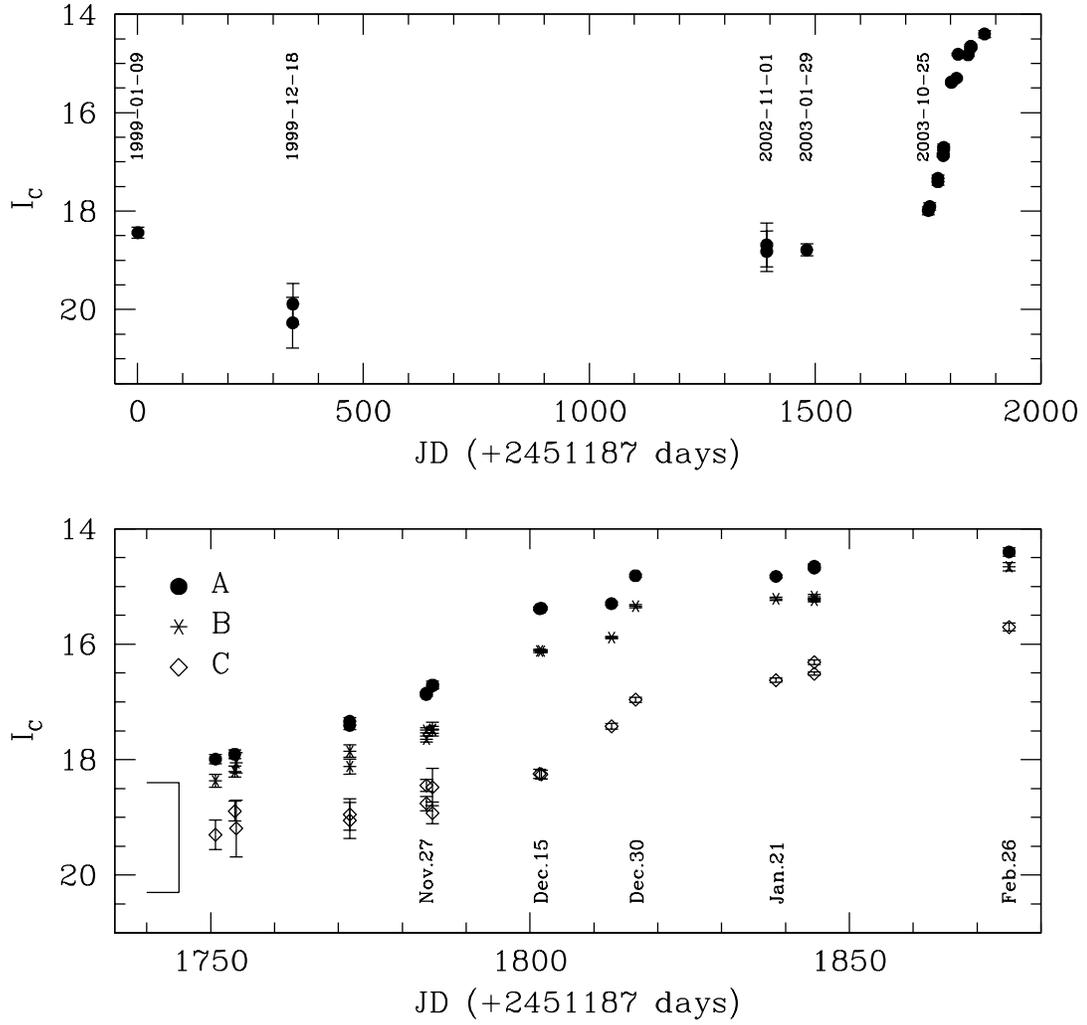}
\caption{Light curve (I-band) of the outburst. In the upper panel we show the
data for source A (Fig. 2), from Jan.09, 1999 to Feb. 26, 2004. 
In the lower panel we show an expanded view of time during the outburst. 
The vertical bar indicates the brightness range during pre-outburst.}
\end{figure}

\begin{figure}
\plotfiddle{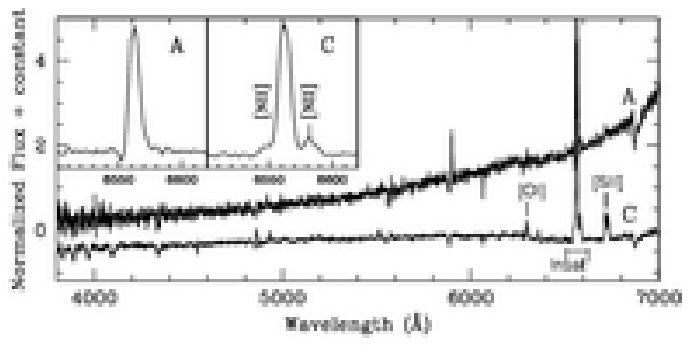}{0in}{0}{500.0}{500.0}{-30.0}{300.0}
\vskip -10.0cm
\plotfiddle{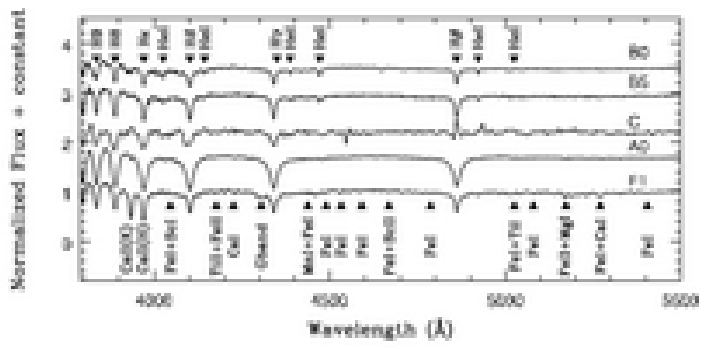}{0in}{0}{500.0}{500.0}{-30.0}{300.0}
\vskip -10.0cm
\caption{FAST Spectra. In the upper panel we show the spectra of sources
A (the exciting star) and C (at the location of HH 22), 
and a detail of the region around H$\alpha$ in the inset. 
The lower panel contains an expanded view of the blue end of the spectrum
of C, showing the higher Balmer lines clearly in absorption. For 
comparison purposes we show spectra of several standard stars.}
\end{figure}

\clearpage

\begin{deluxetable}{cccccc}
\tabletypesize{\scriptsize}
\tablecolumns{6}
\tablewidth{0pc}
\tablecaption{CIDA-Schmidt I-band Photometry}
\tablewidth{0pt}
\tablehead{
\colhead{UT Date} & \colhead{Scan No.} & \colhead{UT} & \colhead{JD} & 
\multicolumn{2}{c}{Source A} \\
\cline{5-6} \\
\colhead{yyyy mm dd} & \colhead{} & \colhead{h:mm:ss} & \colhead{} & 
\colhead{$\rm{I_C}$} & \colhead{$\sigma$($\rm{I_C}$)}
}
\startdata
1999  01 09  & 528 &   6:18:43 &  2451187.7630  & 18.44 & 0.11   \\
1999  12 18  & 503 &   4:10:58 &  2451530.6743  & 20.27 & 0.52   \\
1999  12 18  & 504 &   6:39:32 &  2451530.7775  & 19.89 & 0.42   \\
2002  11 01  & 503 &   8:05:18 &  2452579.8370  & 18.69 & 0.45   \\
2002  11 01  & 504 &   9:35:02 &  2452579.8993  & 18.82 & 0.41   \\
2003  01 29  & 511 &   0:38:00 &  2452668.5264  & 18.79 & 0.12   \\
2003  10 25  & 550 &   5:26:15 &  2452937.7266  & 17.99 & 0.08   \\
2003  10 28  & 501 &   8:54:00 &  2452940.8708  & 17.91 & 0.08   \\
2003  11 15  & 500 &   5:46:19 &  2452958.7405  & 17.34 & 0.07   \\
2003  11 15  & 501 &   6:28:19 &  2452958.7697  & 17.40 & 0.07   \\
2003  11 27  & 501 &   4:48:12 &  2452970.7001  & 16.85 & 0.03   \\
2003  11 27  & 502 &   5:31:12 &  2452970.7300  & 16.87 & 0.03   \\
2003  11 28  & 511 &   4:48:12 &  2452971.7001  & 16.71 & 0.07   \\
2003  11 28  & 512 &   5:34:12 &  2452971.7321  & 16.72 & 0.04   \\
2003  12 15  & 501 &   4:40:00 &  2452988.6944  & 15.38 & 0.02   \\
2003  12 15  & 502 &   5:26:34 &  2452988.7268  & 15.38 & 0.02   \\
2003  12 26  & 500 &   5:22:04 &  2452999.7237  & 15.30 & 0.02   \\
2003  12 30  & 501 &   3:44:00 &  2453003.6556  & 14.81 & 0.03   \\
2004  01 21  & 550 &   2:00:00 &  2453025.5833  & 14.83 & 0.02   \\
2004  01 27  & 551 &   3:26:50 &  2453031.6436  & 14.68 & 0.03   \\
2004  01 27  & 552 &   4:10:00 &  2453031.6736  & 14.65 & 0.04   \\
2004  02 26\tablenotemark{a}  & \nodata & 3:06:00 & 2453061.9208 &  14.40 & 0.08  \\
\enddata
\tablenotetext{a}{FLWO 1.2m + 4Shooter CCD Mosaic camera observation. 
Combination of four 60s exposures.} 
\end{deluxetable}

\end{document}